\documentstyle[12pt,twoside]{article}
\begin{document}
\begin{center}

{\Large\bf IN THE SEARCH OF
SINGULARITY-FREE\\[5PT] 
COSMOLOGICAL MODELS\\[5PT]
IN EFFECTIVE ACTIONS\\[5PT]}
\medskip
 
{\bf
J. C. Fabris\footnote{e-mail: fabris@cce.ufes.br},}  \medskip

Departamento de F\'{\i}sica, Universidade Federal do Esp\'{\i}rito Santo, 
CEP29060-900, Vit\'oria, Esp\'{\i}rito Santo, Brazil \medskip

\end{center}
 
\begin{abstract}
\vspace{0.7cm}
Fundamental theories, like strings, supergravity, Kaluza-Klein, lead after
dimensional reduction and a suitable choice of field configurations, to
an effective action in four dimensions where gravity is coupled non-mininally
to one scalar field, and minimally to another scalar field. These scalar fields
couple non-trivially between themselves. A radiation field is also considered in
this effective action. All the possibilities connected to those fundamental theories
are labeled by two parameters $n$ (related to the non-trivial coupling of the scalar
fields) and $\omega$, connected to the coupling of the Brans-Dicke like field to gravity.
Exact solutions are found, exhibiting a singulariy-free behaviour,
from the four dimensional point of view, for some values
of those parameter. The flatness and horizon problems for these solutions are also analyzed.
It is discussed to which extent the solutions found are non-singular from the point of view of the original frames.
This reveals to be a much complex problem.

PACS number(s): 98.80.Bp, 98.65.Dx
\end{abstract}

\section{Introduction}

The standard cosmological model exhibits some very important achievements in
the description of the Universe. The survey of galaxies and clusters of galaxies
has supported the hypothesis of homogeneity and isotropy in scales
larger than $100\,Mpc$. The discovery of the cosmic microwave background
radiation in 1965 has confirmed the existence of a hot, radiative phase which ended up
at about $z = 1.100$ ($t \sim 10^{12}\, s$). On the other hand, the primordial abundance
of light elements has tested this model up to $z = 10^8$ ($t = 1\, s$) with an impressive accuracy. There are claims
that the spectrum of the anisotropy detected in the cosmic microwave background radiation
tests inflation, creating a window to the deep primordial Universe ($t \sim 10^{-33}$).
But there are many controversies about this question, and for the moment we can not put
the informations obtained from the anisotropy of $CMBR$ as the most remote test of the standard
model. Anyway, the standard model has a very good health and it furnishes a very adequate
description of the Universe in large scale (for a detailed description of the standard model \cite{peacock}).
\par
However, the standard cosmological model has a major drawback: there is an initial singularity.
Since a singularity, where all physical quantities assume an infinite value, is physically
unacceptable, the standard model must be modified, at least at very large energy scales,
in order to give place to a complete regular model. The problem is how to achieve this regular
model through reasonable assumptions. In any case, to avoid the initial singularity
the energy conditions must be violated when the Universe approaches this singularity.
\par
We can think essentially in two scenarios where the singularity can be avoided.
In the first one, the Universe "borns" already in a non-singular state. This idea
has been developed first by Zeldovich and Sakharov (for a description of
this mechanism, see \cite{zel}). The birth of the Universe can occur due to a
fluctuation of a quantum vacuum state. In this case, we can still speak on an age of
the Universe, since a finite time has ellapsed since the phenomena of creation.
Other possibility is that the Universe has ever existed, and it has evolved
in such a way that the singularity had never occured. Some dynamical process must
generate this eternal Universe, in general by the coupling of gravity to some
fundamental fields, which must receive their justification from a fundamental physical
theory.
\par
In this work, we will be interested mainly in the second possibility. Pioneer works
in this sense were done by Melnikov and Orlov \cite{melnikov}, and by Novello, Salim and
Heintzmann \cite{novello1,novello2}. In \cite{melnikov}, a quantum scalar field was considered, leading
to an effective energy-momentum tensor, while in \cite{novello1,novello2} a non-minimal coupling between gravity
and the electromagnetic tensor were studied. In both cases, the behaviour of the scale factor is given by the
expression
\begin{equation}
\label{eu}
a(t) = a_0\sqrt{t^2 + C^2} \quad ,
\end{equation}
where $t$ is the cosmic time and $C$ is a constant.
The solution (\ref{eu}) describes an eternal Universe, without any singularity in the course of its
evolution. The horizon distance is infinite at $t = - \infty$. Hence, no horizon problem appears in
these kind of model. The flatness problem, as well as the structure formation problem, are more
delicate points, which deserves a deeper investigation. In any case, these kind of cosmological models
constitute a very interesting and appealing alternatives to the standard cosmological scenario with
an initial singularity which, anyway, may be eliminate in a way or other.
\par
In what follows, we will describe how similar models can appear in effective action obtained from
fundamental theories, like superstring, supergravity or Kaluza-Klein.
In their scalar-tensor sector, in the lower energy limit, these fundamental theories
predict a non-minimal coupling of gravity and a scalar field (a "Brans-Dicke"-like term),
which can be
a dilaton field for string theories, or the field connected with compactified dimensions
in the case of theories formulated in a higher dimensional space-time (what does not
exclude
the string case).
Moreover, others scalar field appear, and in general they are associated with
some components of gauge fields in the multiplet of those fundamental theories.
These extra scalar fields may couple non-trivially with the Brans-Dicke field,
in a way which depends on the original theory which gives birth to
the effective action.
\par
The different possible origins
of the effective action will be labeled by two parameters, $\omega$ and $n$, connected with how a
Brans-Dicke type field couples to gravity and to others scalar fields that appear in the effective
action. In the study that will be carried out in the present work, we will consider that this effective action is coupled to a radiation field. This is quite natural
since those fundamental fields predict also an electromagnetic field in their vectorial sector.
We will verify to which extent such models may circumvent the singularity problem.
It will be shown that from the four dimensional point of view, there is a class of non-singular cosmological
models.
But, from the point of view of the fundamental theories that motivate the effective action studied here,
a singularity remains, with the remarkable possible exception of the string theory, due to its
duality properties. This reveals the difficulties to construct a true regular cosmological model.
Some other consequences, like the flatness and horizon problems, will be analyzed.
Remarkably, at least the horizon problem may be solved only for the (four dimensional) singularity-free scenarios.
\par
In the next section, we will describe the effective action and its connection to those fundamental theories.
In section three, the cosmological solutions which come from it will be settled out, and their general behaviour
will be classified. In section four, the consequences for the flatness and horizon problems will be
analyzed. In section five, general conclusions about these models will be sketched.

\section{Effective actions in four dimensions}

Superstring, Supergravity and Kaluza-Klein theories predict, at energy levels much smaller than the
Planck energy, the coupling of gravity to scalar and gauge fields. When studying cosmological scenario,
it must be retained in these effective action the fields of spin zero and two (scalar and gravitational fields).
This can be achieved in the dimensional reduction process since those theories are generally formulated in
a higher dimensional space-time.
The general effective action in four dimensions can be written as
\begin{equation}
\label{action}
L = \sqrt{-g}\biggr(\phi R - \omega\frac{\phi_{;\rho}\phi^{;\rho}}{\phi} - \phi^{n}\Psi_{;\rho}\Psi^{;\rho}
-\chi_{;\rho}\chi^{;\rho}\biggl)
+ L_m
\end{equation}
where $\phi$ is a dilaton (or Brans-Dicke) type field, $\Psi$ is a scalar component (in four dimensions) which comes
from the gauge fields in the original multidimensional space-time. Scalar fields coming from the Ramond-Ramond sector of supersting can also be taken into
account, and they are represented by the field $\chi$.
$L_m$ is the lagrangian density of some
other matter components. We will be interested mainly, in what follows, in the radiation field, which can
be regarded as a averaged component of the electromagnetic field which is also present in the multiplet of
those fundamental theories.
\par
The Lagrangian (\ref{action}) describes an effective action which comes from many different
fundamental theories. In order to be specific, we may quote some special cases:
\begin{enumerate}
\item It corresponds to pure multidimensional theories with $\Psi = \chi =$
constant and $\omega = (1 - d)/d$, where $d$ is the 
number of compactified dimensions (in this case, we consider the
compactification on the torus); 
\item If we consider a two form-gauge field in higher dimensions
we obtain $\Psi \neq$ constant, and $n = - \frac{2}{d} + 1$;
\item A conformal gauge field, represented by a $(d + 4)/2$-form, leads to
$n = - \frac{2}{d}$ \cite{fabris}; 
\item In string theory $\omega = -1$; moreover, the fact that in general
the three-form field $H_{\mu\nu\lambda}$ has, in 
four dimensions,
just one degree of freedom permits one to write $H^{\mu\nu\lambda} =
\phi^{-1}\epsilon^{\mu\nu\lambda\sigma}\Psi_{;\sigma}$
which corresponds to $n = - 1$ in (\ref{action}) \cite{maeda}.
\end{enumerate}
Of course, (\ref{action}) does not cover all possible cases, with the many possible field configurations
and compactification scheme. But, some general features are covered by it.
\par
Many interesting studies have already been carried out for some particular cases of
(\ref{action}). One of the pioneer works in this sense
has been done in \cite{freund}, where cosmological scenarios emerging from eleven and ten dimensional
supergravity theories have been outlined. In this work, connections with the Brans-Dicke theory had
already been established, although all the generality of the Brans-Dicke parameter $\omega$ is not
covered by those fundamental theories as, by the way, by any other quoted before. The cosmological
scenarios settled out in \cite{freund} present all an initial singularity. But, this work had the advantage of
calling attention that supergravity theories had important consequences not only for particle physics but also
for cosmology, suggesting, in another way, the deep connection between these two extreme (in what concerns
length scales) situations.
\par
In what concerns string cosmology, there is a large literature about this subject. In fact, it seems
that cosmology has become the only arena where the string idea can be tested. In principle, the specific
effects coming from strings manifestate at an energy level of about Planck energy, hence too far from the
terrestrial avaliable accelerators. Then, it is possible to test this theory only through some traces
left by a "string" phase in the very early Universe.
A very popular cosmological scenario based on string theory is the so-called {\it pre-big bang cosmology}
(to a comprehensive review, see \cite{gasperini}). In the pre-big bang scenario, there is a curvature growing
phase prior to the decreasing curvature phase of the FLRW Universe. In order to construct a complete scenario,
the dualities properties of the string effective action at tree level, without the axion term, are explicit
employed. However, the transition from the curvature growing phase to the decreasing curvature phase,
which must be achieved by considering other corrections to the string effective action as
the curvature reachs very high values, is not very well understood until now, and we may consider that the
pre-big scenario is still a program under development, with many unsolved questions.
\par
However, string cosmology does not end with the pre-big bang model. There is a lot of work in cosmology
employing the multiplet predicted, at lower energy level, by string action. A very exhaustive
description of the status of string cosmology is given in \cite{copeland}. One important point in
considering string cosmology is that besides the well-known problem of the curvature singularity, that
plagues the standard cosmological scenario, it appears an expansion parameter singularity: if the dilaton
field goes to zero, the string expansion parameter may diverge and the effective action makes no sense anymore.
It is generally argued that such new divergence may be coped with through the dualities properties of
string theory: due to the S-duality, the strong coupling regime may be mapped into a weak coupling regime of
another formulation of string theory \cite{polchinski}. In practice, it becomes very difficult to show that this scheme works
since the analysis of the S-duality in a non-trivial background, like the geometry of a FLRW Universe, is
a very difficult task.
\par
Concerning yet the possible values of $\omega$ in (\ref{action}), it must be stressed that
p-branes configurations in string models, may lead to a more general expression for this
coupling parameter. Explicitly, a p-brane,
a $p$ dimensional extended object, in $d$ dimensions leads to \cite{duff}
\begin{equation}
\omega = - \frac{(d-1)(p-1) - (p+1)^2}{(d-2)(p-1) - (p+1)^2}
\end{equation}
Hence, $\omega = - 1$ is just one special case for a one-brane, that is, a pure string
\cite{kirill}.
\par
Even if special attention was given for the connection between (\ref{action}) to the string
effective action in lower energies, it must be kept in mind that (\ref{action}) covers
a quite large number of possible effective actions obtained from fundamental theories, as
those quoted before. However, it does not cover possible generalizations of the Einstein-Hilbert
action, like that proposed by \cite{novello1,novello2}, and the cosmological scenarios that
can be obtained from (\ref{action}) have at same time strinking similarities and differences
with that represented by the solution (\ref{eu}), as it will be seen later.

\section{Cosmological solutions}

Let us concentrate ourselves now on the search of exact cosmological solutions in
the realm of the effective model described before.
The case with Ramond-Ramond field $\chi$, with no phenomenological matter, has been studied in \cite{picco}.
We will not consider this field anymore during the present work.
The field equations corresponding to the Lagrangian (\ref{action}) are
\begin{eqnarray}
\label{fe1}
R_{\mu\nu} - \frac{1}{2}g_{\mu\nu}R &=& \frac{8\pi}{\phi}T_{\mu\nu} +
\frac{\omega}{\phi^2}\biggr(\phi_{;\mu}\phi_{;\nu}
- \frac{1}{2}g_{\mu\nu}\phi_{;\rho}\phi^{;\rho}\biggl) +
\frac{1}{\phi}\biggr(\phi_{;\mu;\nu} - g_{\mu\nu}\Box\phi\biggl) + \nonumber\\
&+& \phi^{n-1}\biggr(\Psi_{;\mu}\Psi_{;\nu} -
\frac{1}{2}g_{\mu\nu}\Psi_{;\rho}\Psi^{;\rho}\biggl) 
\quad ;\\
\nonumber\\
\label{fe2}
\Box\phi &+& \frac{1 - n}{3 + 2\omega}\phi^n\Psi_{;\rho}\Psi^{;\rho} = \frac{8\pi}{3 + 2\omega}T
\quad ; \\ 
\nonumber \\
\label{fe3}
\Box\Psi + n\frac{\phi_{;\rho}}{\phi}\Psi^{;\rho} &=& 0 \quad ; \\
{T^{\mu\nu}}_{;\mu} &=& 0 \quad .
\end{eqnarray}
The energy-momentum tensor will be that of a perfect fluid:
\begin{equation}
T^{\mu\nu} = (\rho + p)u^\mu u^\nu - pg^{\mu\nu} \quad .
\end{equation}
We will consider that this perfect fluid obeys a barotropic equation of state
$p = \alpha\rho$, where in general $1 \geq \alpha \geq -1$.
The Robertson-Walker metric, describing an isotropic and homogeneous Universe, has the form
\begin{equation}
ds^2 = dt^2 - a^2(t)\biggr[\frac{dr^2}{1 - kr^2} + r^2(d\theta^2 + \sin^2\theta d\phi^2)\biggl] \quad ,
\end{equation}
where $k = 0, -1, +1$ indicates a flat, open or closed spatial section respectively, as usual.
Considering the energy-momentum tensor, the Robertson-Walker metric and the new time variable
definite as $dt = a^3d\theta$, which reveals a quite convenient time reparametrization
when we treat systems involving gravity
and scalar fields, the equations of motion take the form
\begin{eqnarray}
\label{em1}
3\biggr(\frac{a'}{a}\biggl)^2 + 3k a^4 &=& \frac{8\pi}{\phi}\rho a^6 +
\frac{\omega}{2}\biggr(\frac{\phi'}{\phi}\biggl)^2 -
3\frac{a'}{a}\frac{\phi'}{\phi} + 
\frac{\phi^{n-1}}{2}\Psi'^2 \quad , \\
\label{em2}
\phi'' + \frac{1 - n}{3 + 2\omega}\phi^n\Psi'^2 &=& 
\frac{8\pi}{\phi}(\rho - 3p)a^6 , \quad , \\
\label{em3}
\Psi'' + n\frac{\phi'}{\phi}\Psi' &=& 0 \quad , \\
\label{em4}
\rho' + 3\frac{a'}{a}(\rho + p) &=& 0 \quad ,
\end{eqnarray}
where the primes mean derivatives with respect to $\theta$.
\par
The equations (\ref{em3}) and (\ref{em4}) have simple first integrals:
\begin{equation}
\Psi' = A\phi^{-n} \quad , \quad \rho = 3Ma^{-3(1 + \alpha)}
\quad ,
\end{equation}
where $A$ and $M$ are integration constants.
\par
Now, we specialize our analysis to the case $p = - \frac{\rho}{3}$, that is, a radiative
fluid. This case is very important, first because a radiative fluid plays a crucial role
in the primordial Universe, and second because an eletromagnetic field coupled to gravity
is a general prediction of those fundamental theories quoted in the previous section that
motivate (\ref{action}).
\par
Inserting the equation of state of a radiative fluid, equation (\ref{em2})
becomes homogeneous and admits a simple first integral:
\begin{equation}
\label{nem2}
\frac{\phi'^2}{2} + \frac{A^2}{3 + 2\omega}\phi^{1-n} = \frac{D}{2} \quad .
\end{equation}
This first order non-linear equation can be explicitly solved for any value of $\omega$ and $n$
through a new time parametrization. In fact, defining
\begin{equation}
d\theta = \sin^\frac{1 + n}{1 - n}\xi d\xi
\end{equation}
equation (\ref{nem2}) admits the general solution
\begin{equation}
\phi(\xi) = \biggr(\frac{(3 + 2\omega)D}{2A^2}\biggl)^{2/(1 - n)}\sin^{2/(1 - n)}\xi \quad .
\end{equation}
\par
We turn now to the integration of equation (\ref{em1}), in order to obtain the behaviour of
the scale factor. To do this, it reveals convenient to redefine the scale factor
as $a = \phi^{-1/2}b$. This is equivalent to perform a conformal transformation in
(\ref{action}), rewritting it in the so-called Einstein's frame. However, we consider this transformation
just as a procedure to make the equation easier to integrate, and we do not attribute to it any deeper
meaning. In other words, the physical frame is the Jordan's frame, with a non-minimal coupled scalar field.
With that transformation, equation (\ref{em1}) takes the form,
\begin{equation}
\biggr(\frac{b'}{b}\biggl)^2 = \frac{(3+2\omega)D}{12\phi^2}\biggr[1 + rb^2 + ks^2b\biggl] \quad ,
\end{equation}
where $r = \frac{12M}{3 + 2\omega}$ and $s^2 = \frac{12}{(3+2\omega)D}$.
This expression can be recast in an integral form:
\begin{equation}
\int \frac{db}{b\sqrt{1 + rb^2 - ks^2b^4}} = \pm \sqrt{\frac{(3+2\omega)D}{12}}\int \frac{d\theta}{\phi}
\quad .
\end{equation}
Transforming this relation using the new time parameter $\xi$ and the solution already found for
the field $\phi$, we can obtain explicit solutions for $a$. This depends, of course,  on
the value of $k$, the curvature of the spatial section.
We present in what follows separately the final expressions, where in all cases
$p = \pm \frac{1}{1 - n}\sqrt{1 + \frac{2}{3}\omega}$ and $R = 2s/r$.

\begin{itemize}
\item $k = 1$:
\begin{equation}
\label{s1}
a(\xi) = a_0\sin^{1/(n-1)}\xi\sqrt{1 + \frac{2RCg(\xi)}{1 + g^2(\xi)}} \quad ,
\end{equation}
where
\begin{equation}
g(\xi) = - RC + f(\xi) \quad , \quad f(\xi) = R\frac{1 + c^2\tan^{2p}(\frac{\xi}{2})}{1 - c^2\tan^{2p}(\frac{\xi}{2})}
\quad , \quad  C = \sqrt{1 + 1/R^2} \quad ,
\end{equation}
$c$ being another integration constant.
\item $k = 0$:
\begin{eqnarray}
\label{s2}
a = a_0(\sin\xi)^{1/(n-1)}\xi\biggr\{\frac{\tan^{p}(\xi/2)}{1 -
c^2\tan^{2p}(\xi/2)}\biggl\} \quad .
\end{eqnarray}
\item $k = - 1$:
\begin{itemize}
\item $R > 1$:
\begin{equation}
\label{s3}
a(\xi) = a_0\sin^{1/(n-1)}\xi\sqrt{- 1 + \frac{2RCg(\xi)}{1 - g^2(\xi)}} \quad ,
\end{equation}
where now
\begin{equation}
g(\xi) = - RC + f(\xi) \quad , \quad f(\xi) = R\frac{1 + c^2\tan^{2p}(\frac{\xi}{2})}{1 - c^2\tan^{2p}(\frac{\xi}{2})}
\quad , \quad C = \sqrt{1 - \frac{1}{R}};
\end{equation}
\item $R = 1$:
\begin{equation}
\label{s4}
a(\xi) = \sin^{1/(n - 1)}\xi\sqrt{\frac{c^2\tan^{2p}(\frac{\xi}{2})}{1 - c^2\tan^{2p}(\frac{\xi}{2})}} \quad .
\end{equation}
\item $R < 1$:
\begin{equation}
\label{s5}
a(x) = a_0\sin^{1/(n - 1)}\xi\sqrt{-1 + RC\frac{1 + f^2(\xi)}{1 - f^2(\xi)}} \quad ,
\end{equation}
where now
\begin{equation}
\quad f(\xi) = \frac{1}{EC}\frac{1 + c^2\tan^{2p}(\frac{\xi}{2})}{1 - c^2\tan^{2p}(\frac{\xi}{2})}
\quad , \quad E = 1/RC + 1 \quad , \quad C = \sqrt{- 1 + \frac{1}{R}} \quad .
\end{equation}
\end{itemize}
\end{itemize}
\par
The solutions described before are not valid for $n = 1$. For this special case, we have the following
solution for the "Brans-Dicke" scalar field:
\begin{equation}
\phi = L^2\theta \quad , \quad L^2 = D - \frac{2A^2}{3+2\omega} \quad .
\end{equation}
Defining $B = \sqrt{\frac{(3+2\omega)D}{12L^2}}$, the equations can be integrated in terms
of the time parameter $\theta$.
We find that the scale factor behave as
\begin{itemize}
\item $k = 1$:
\begin{equation}
a(\theta) = a_0\frac{1}{\sqrt{\theta}}\sqrt{1 + \frac{2RCg(\theta)}{1 + g^2(\theta)}} \quad ,
\end{equation}
where
\begin{equation}
g(\theta) = - RC + f(\theta) \quad , \quad f(\theta) = R\frac{1 + c^2\theta^{2B}}{1 - c^2\theta^{2B}}
\quad . 
\end{equation}
The parameter $C$ is defined as in the case $k = 1$, $n \neq 1$.
\item $k = 0$:
\begin{equation}
a(\theta) = a_0\frac{1}{\sqrt{\theta}}\frac{1}{(- \theta^B + \theta^{-B})} \quad .
\end{equation}
\item $k = - 1$:
\begin{itemize}
\item $R > 1$:
\begin{equation}
a(\theta) = a_0
\frac{1}{\sqrt{\theta}}\sqrt{- 1 + \frac{2RCg(\theta)}{1 - g^2(\theta)}} \quad ,
\end{equation}
where now
\begin{equation}
g(\theta) = - RC + f(\theta) \quad , \quad f(\theta) = R\frac{1 + c^2\theta^{2B}}{1 - c^2\theta^{2B}}
\quad ,
\end{equation}
with $C = \sqrt{1 - \frac{1}{R}}$;
\item $R = 1$:
\begin{equation}
a(\theta) = a_0\frac{1}{\sqrt{\theta}}\sqrt{\frac{c^2\theta^{2B}}{1 - c^2\theta^{2B}}} \quad .
\end{equation}
\item $R < 1$:
\begin{equation}
a(\theta) = a_0\frac{1}{\sqrt{\theta}}\sqrt{\frac{2RCf(\theta)}{1 - f^2(\theta)} - 1} \quad ,
\end{equation}
where
\begin{equation}
\quad f(\theta) = \frac{1}{EC}\frac{1 + c^2\theta^{2B}}{1 - c^2\theta^{2B}}
\quad , \quad E = 1/RC + 1 \quad ,
\end{equation}
with $C = \sqrt{- 1 + \frac{1}{R}}$.
\end{itemize}
\end{itemize}
\par
The general features of the solutions found before are very similar to those found
in \cite{picco}. In spite of the complexity of the solutions exhibited above, an asymptotic
analysis, as it will be done in the next section in more details, permit to sketch the general
behaviour of the solutions for the various possible cases. When $n < 1$ and $\omega < 0$, the solutions exhibit a bounce,
the scale factor having an initial and final infinite value.
However, these scenario are free from curvature singularities only if $\omega < - 4/3$, and the time variable is
definite in the interval $- \infty < t < \infty$. Hence, in these
cases, $\xi \rightarrow 0$ implies $t \rightarrow - \infty$. For $n < 1$ and $- 4/3 < \omega < 0$,
there is an initial singularity as $\xi \rightarrow 0$, in spite of the fact that the model exhibits a
bounce. The time variable is definite in the interval $0 < t < \infty$.
If $n > 1$ or $\omega > 0$,
the solutions present also an initial singularity. Moreover, the scale factor goes to zero as $\xi \rightarrow 0$,
and the scenario is very similar to the standard one. The same
behaviour occurs in the particular case
$n = 1$.

\section{Asymptotical behaviour: the flatness and horizon problems.
New considerations on the singularity problem}

In what follows, we will concentrate ourselves on the solutions for which $n \neq 1$.
As it has been stressed before, in spite of the complexity of the solutions determined before, the asymptotical analysis
reveals a very simple behaviour in both extremity of the allowable values of $\xi$, the
new time parameter employed in the derivation of those solutions. We will detail now some aspects of
this asymptotical analysis. In this discussion we will consider $p > 0$, since the cases with
$p < 0$ imply just to reverse the time direction.
Let us consider first the asymptotic behaviour as $\xi \rightarrow 0$. In this case,
all the solutions exhibited before lead to the same expression, independently of
the curvature:
\begin{equation}
\label{fa}
a(\xi)\sim \xi^{\frac{1}{1 - n}(-1 + \sqrt{1 + \frac{2}{3}\omega})} \quad .
\end{equation}
In terms of the cosmic time, (\ref{fa}) may be expressed as
\begin{equation}
a(t) \sim t^\frac{1 + \omega - \sqrt{1 + \frac{2}{3}\omega}}{4 + 3\omega} \quad,
\end{equation}
which is the same expression that occurs for the Brans-Dicke vacuum solutions.
The scalar field $\phi$ exhibits also a behaviour typical of the Brans-Dicke
vacuum:
\begin{equation}
\phi \propto t^\frac{1 - 3\sqrt{1 + \frac{2}{3}\omega}}{4 + 3\omega} \quad .
\end{equation}
For $\omega > - 4/3$, $\xi \rightarrow 0$ means $t \rightarrow 0$, while
for $\omega < - 4/3$, it implies $t \rightarrow - \infty$.
\par
In the other asymptotic, the expression of the scale factor depends on the curvature of
the spatial section:
\begin{itemize}
\item $k = 0$:
\begin{equation}
\label{fa1}
a \propto t^{1/2} \quad , \quad \phi = \mbox{constant} \quad ;
\end{equation}
\label{fa2}
\item $k = \pm 1$:
\begin{equation}
a \propto t^\frac{1 + \omega + \sqrt{1 + \frac{2}{3}\omega}}{4 + 3\omega} \quad ,
\quad \phi \propto t^\frac{1 + 3\sqrt{1 + \frac{2}{3}\omega}}{4 + 3\omega} \quad .
\end{equation}
\end{itemize}
There is one exception for the last case: $k = - 1$ and $R = 1$. For this special value of the parameter
$R$, the asymptotic behaviour reads
\begin{equation}
a(t) \propto t \quad , \quad \phi(t) \propto \mbox{constant} \quad .
\end{equation}
It means that the Minkowski space is recovered asymptotically.
\par
Now, we are in position of analyzing the horizon and the flatness problems in the
context of the models exposed before.
In what concerns the horizon problem,
we must evaluate the expression for the proper horizon distance:
\begin{equation}
d_h = a(t)\int_{t_i}^t\frac{dt'}{a(t')} \quad ,
\end{equation}
where $t_i$ is the initial time. For $\omega > - 4/3$, this initial time occurs for $t = 0$, while for
$\omega < - 4/3$, it occurs for $t \rightarrow - \infty$. Hence, in the first case, there is a horizon,
as in the standard model, and we must expect all the problems due to this fact, including those refering
to the formation of structure. In the second case, the horizon diverges initially; consequently, all the
space-time is causally connected from the begining. No horizon problem occurs as it happens with
the models developed in \cite{novello1,novello2}. It is remarkable that the solution for the horizon is
verified in the models where no curvature singularity occurs.
\par
For the flatness case, the situation is somehow curious. From expressions (\ref{s1}-\ref{s5}),
we can verify that the asymptotic $t \rightarrow \infty$ occurs for $\xi \sim \pi/2$ when $k = 0$,
and for $\xi \sim \pi$, for $k = \pm 1$ (with the exception of the particular case
$k = - 1$ and $R = 1$, which we will ignore in what follows). This leads to complete diferent asymptotic expressions for the
scale factor. For $k = 0$, the asymptotic behaviour coincides with the radiative flat standard model, while
for $k = \pm 1$,
we verify that the solutions go asymptotically to
the Brans-Dicke vacuum flat
solutions. Hence, in principle the traces of the curvature are washed out during the evolution of
the Universe. But, the asymptotic behaviour depends on the curvature term $k$ even
if it exhibits always a flat-like solution. So, in one sense it can be argued that traces of the curvature
are not absent in the asymptotical flat behaviour.
\par
The case $k = 0$ presents also the interesting feature that the dilatonic scalar field goes asymptotically to
a constant value. We can fix this value by identifying its inverse to the (constant) gravitational coupling.
Hence, this particular solution opens the possibility to give a precise value to the dilatonic field, and hence
it may give
some clues on the phenomenology of string theories.
\par
The asymptotical behaviour described before, reveals a class of solutions which are free from
curvature singularity. It means that, the invariants of curvature are regular during all the evolution of the
Universe when $\omega < - 4/3$. This can be easily seen analyzing, for example, the Ricci scalar, $R$.
In the asymptotical regions, this term behaves as
\begin{equation}
R \propto t^{-2} \quad .
\end{equation}
If the asymptotics occur for $t \rightarrow \pm \infty$, then $R \rightarrow 0$. As the scale factor never goes to zero,
than the solution is complete regular in this sense.
This happens, as it has already be remarked, for $\omega < - 4/3$. For all other cases, the Universe begins at
$t = 0$ and the invariants of curvature diverge: the geometry exhibits a singularity.
\par
However, we must remember that the effective action studied above is connected with string or multidimensional
theories. In this case, the analysis of singularity is more delicate. In fact, in the case of multidimensional
theories, the scalar field $\phi$ is connected with the scale factor of the internal dimensions, and if it
goes to zero, the invariants of curvature written in the higher dimensional original space-time diverge.
So, even if the effective model in four dimension is regular, the original theory presents a breakdown of
the geometry. This happens for all cases studied previously. Hence, we can expect the presence of some
singularity if we consider the models studied as coming from a pure multidimensional theory.
\par
The situation is more involved if these models come from string theory. In this case, the
effective model analyzed has been compactified in the torus, and no dynamics is attributed to
the internal dimensions. On the other hand, the scalar field $\phi$ is nothing more than the dilaton field,
which is conneceted with the string expansion parameter by the expression $g_s = \phi^{-1}$.
This expansion parameter controls the perturbative series which gives birth to the effective
action (\ref{action}). If $\phi$ goes to zero, as it actually happens when $\xi \rightarrow 0$, then $g_s \rightarrow \infty$,
and this perturbative series, in principle, makes no sense anymore.
\par
However, string theories, as it has already been stated, have the so-called T-duality and S-duality.
The T-duality connects theories with a small radius in the internal dimension with theories with a large
radius in the internal dimension. On its side, S-duality connects theories in the strong coupling regime,
as it happens for large values of $g_s$, with theories in the weak strong regime, where the effective
action employed here makes sense. A problem that arises is due to the fact that mapping a string theory in another string
theory, the field multiplet is different, mainly for the Ramond-Ramond sector. But, for some cases,
as in the type IIB string theory, the S-duality connects the theory with itself. Hence,
the singularity in the string expansion parameter may be avoided without essentialy touching the
effective model in the lower energy limit. The big problem concerning this point is to prove that this happens for the
cosmological models studied previously, where the highly non-trivial background space-time makes such
a study a very hard task.

\section{Conclusions}

The standard cosmological model has very impressive features, being compatible with
the avaliable observations data, specially those concerning the primordial nucleosynthesis,
measurements of the cosmic microwave background radiation, Hubble expansion, galatic survey, etc.
However, the extrapolation of this model to very primordial period indicates the existence of
an initial singularity. This is, of course, a major drawback for this model, since no
physics can be applied to a singular state. Moreover, very special initial conditions are
required by the standard cosmological model in order to obtain those agreements between observations
and theory.
\par
The question of the particular initial conditions required by the standard cosmological model
may be coped with by the inflationary scenario, in which a brief phase of accelerated expansion
prior to the nucleosynthesis era is implemented. Even if the inflationary scenario is make possible
by the use of phenomenological potentials for the inflaton field (the field responsible for this
accelerated expansion), its achievemnts in order to solve those initial conditions problems
makes it a widely accepted mechanism for obtaining a model of the Universe with the desired features, in particular
homogeneous and isotropic.
\par
However, the initial singularity problem remains a serious obstacle to construct a complete cosmological
model valid for any time. It is generally argued that quantum effects may give a positive answer to this
problem. But, until now no consistent quantum theory of gravity is avaliable; hence this possibility
remains a pure hope.
On the other hand, some classical attempts have being tried in order to obtain a complete cosmological
scenario, free from singularities problems.
One remarkable example is the eternal solution found in \cite{novello1,novello2}, where gravity
is coupled non-minimally to an electromagnetic field. This solution is complete free from singularities and,
at same time, it has no horizon problem.
\par
In this work we tried to answer the question if a singularity-free solution can be obtained from
theories that may be considered as fundamental: string, supergravity, Kaluza-Klein, etc.
These theories predicted that gravity is coupled to other fields, scalar or gauge fields, in general
in a higher dimension space-time. When we perform a dimensional reduction, and specific configurations
for the various field are chosen, which can be of cosmological interest, an effective action in four
dimension is obtained. We have considered a quite simple
version of this effective action which is labeled by two parameters: $n$, which is linked to
the coupling of the scalar field between themselves; and $\omega$, which determines the coupling of
a dilaton-like field to gravity and matter. The different possible values of these two paramenters
permit to trace the connection of the effective action with one of the possible original theories.
Since an electromagnetic field is always present in the multiplet of the original theories,
we have coupled the effective action to a radiation field, represented by a fluid with
an equation of state $p = \frac{\rho}{3}$.
\par
Exact solutions for the resulting model were found. Essentially, they describe a bouncing
Universe if $n < 1$ and $\omega < 0$. However, such bouncing Universe is non-singular, from the
four dimensional point of view,
only if $\omega < - 4/3$. If $n \geq 1$ or $\omega > 0$, a Friedmann-like Universe occurs, with
an initial singularity, the scale factor having initially a zero value.
An important point is that the cases where $n < 1$ and $\omega < - 4/3$ are
regular only from the four-dimensional point of view. If we remember their muldimensional origin,
in which the field $\phi$ can be seen or as the scale factor of the internal dimensions (multidimensional
theories) or the dilaton field (string theory), a singularity may exist: in the first case, due to
the collapse of the internal dimension; in the second due to divergences in the string expansion parameter,
which is determined by the inverse of the dilaton field.
\par
In the case of string theory, the divergence of the expansion parameter may make the effective action
senseless, since it is essentially a first term of a series, which may not converge if $g_s \rightarrow \infty$. However,
the string theories present important dualities properties. One of the them, the S-duality, connects theory
in the strong coupling regime with theories in the weak coupling regime. This duality may be employed in
order to give meaning to the effective action (and to the resulting models) studied here. But, this
must be explicitly proved, and this is not an easy task due to the non-triviality of the background studied here.
\par
It is remarkable that for the singularity-free models obtained here (from the four dimensional point of view),
the horizon problem is naturally solved: the horizon distance diverges at the initial time. Moreover, it is
possible that the flatness problem may also be solved. However, this is a more delicated point since
the solutions are, for any value of $k$, asymptotically flat, but their exact expression depends on the curvature of the spatial section.
This is a curious point, which deserve a deeper study.
\par
Strictly speaking, we have just established indications that the fundamental theories of physics may lead
to non-singular cosmological scenarios. More investigations are needed to verify if this really happens.
The study carried out here at same time that gives some hope in what concerns the singularity problem,
show how difficult is to treat this feature of the standard cosmological model. In this sense, solutions
as those found in \cite{novello1,novello2} remain a remarkable example of how a complete regular scenario can
be constructed.

\vspace{0.5cm}
{\bf Acknowledgements:} We thank S\'ergio V.B. Gon\c{c}alves for his critical reading of the text and CNPq (Brazil) for partial financial support.

\end{document}